\documentclass[aip,showpacs]{revtex4-1}
\usepackage{graphicx}
\usepackage{amssymb}
\usepackage{amsmath,array}
\usepackage[letterpaper, portrait, margin=3cm]{geometry}
\begin{document}
% Use the \preprint command to place your local institutional report number 
% on the title page in preprint mode.
% Multiple \preprint commands are allowed.
%\preprint{}

\title{Bose condensation and the Casimir effects of an imperfect Bose gas in a d-dimensional configuration space} 

% repeat the \author .. \affiliation  etc. as needed
% \email, \thanks, \homepage, \altaffiliation all apply to the current author.
% Explanatory text should go in the []'s, 
% actual e-mail address or url should go in the {}'s for \email and \homepage.
% Please use the appropriate macro for the type of information

% \affiliation command applies to all authors since the last \affiliation command. 
% The \affiliation command should follow the other information.

\author{G. M. Bhuiyan}
\email[]{gbhuiyan@du.ac.bd}
%\homepage
%\thanks{}
%\altaffiliation{agfds}
\affiliation{Department of Theoretical Physics, University of Dhaka, Dhaka-1000, Bangladesh}
% Collaboration name, if desired (requires use of superscriptaddress option in \documentclass). 
% \noaffiliation is required (may also be used with the \author command).
%\collaboration{}
%\noaffiliation

%\date{\today}

\begin{abstract}
Some properties of an ideal gas of massive bosons in a mean field potential and, confined between two 
infinite parallel slabs in a d-dimensional configuration space are investigated systematically. Here, one particle density of states approach is employed to study the critical temperature, shift of density, Casimir effects and critical exponents,
starting from the evaluation of the grand canonical free energy in d-dimension. We have found that, the shift of density, Casimir force and the critical temperature depend on the space dimensionality. But the Casimir force
decays as an inverse power law of the distance between two slabs in the condensate and, decays exponentially in the non-condensed state situated very close to the point of phase transition. Most importently, this study enabled us to predict the shift of the density of excited  bosons due to mean field potential, and also the dimensional 
dependence of the critical exponent. The form of the critical exponent is found to be $\frac{1}{d-2}$ for the imperfect Bose gas. This leads to a value of critical exponent $1$ for $d=3$. 
\end{abstract}
% insert suggested PACS numbers in braces on next line

\maketitle 
%\maketitle must follow title, authors, abstract and \pacs

% Body of paper goes here. Use proper sectioning commands. 
% References should be done using the \cite, \ref, and \label commands
\section{Introduction}
After the demonstration by Einstein that there is a possibility of condensation ( which is referred to 
as the Bose-Einstein condensation) of the ideal Bose gas (IBG) many attempts have been made to understand this phenomenon. Since then the bulk behavior including condensation characterized by the thermodynamic properties, occupation of the states and their fluctuations are investigated[1-5]. But after the observation of the 
Bose-Einstein Condensation (BEC) of cold atoms in a trap in 1995 [6-8] the interest of better understanding
of Bose condensation has been strongly stimulated. As a result different studies are made on a Bose system analyzing the effects of potentials [9-14], interaction between particles [15-24], space dimensionality [25-30], criticality near the phase transition [26,31,32], and finite particle number [9].

Recently, some attempts have been spent to study imperfect Bose gas (IMBG) plugging in the mean field potential[15,33-35].  Napiorkowski and Piasecki, in [33], evaluated the grand canonical potential considering the 3-dimensional space along with the mean field potential, within the perspective of the Quantum Statistical Mechanics. However, the density of states for this type of imperfect Bose gas has not been studied yet, although it plays an important role in statistical physics [14,25,27]. On the other hand,
theoretical studies demonstrate that the space dimensionality and the type of potential affects different properties such as the density [20], critical temperature [20], the thermodynamic Casimir force and the critical exponents [33]. It is, therefore, interesting to derive one particle density of states in d-dimensional configuration space for the imperfect Bose gas (IMBG) that involves the mean field potential. 
It is also very much interesting to examine then if this density of states approach can predict different physical properties specifically the Bose condensation, density shift and Casimir effects of the IMBG correctly.

In a real system, interaction among the particles does vitally exist, but it makes the problem difficult to solve analytically. In order to make the problem solvable, and to retain the essential physics intact one can  represent a real system approximately by an ideal system of non interacting particles in the presence of a potential or a trap. The validity of the approach is nicely justified by the experimental fact that the influence of interaction between particles on the Bose-Einstein condensation transition  temperature is only of several percent [16].

In this report, we intend to derive first the one particle density of states for the ideal Bose gas in a mean field  potential (MFP) using a d-dimensional configuration space and then to apply it to evaluate the grand potential energy for a system confined between two parallel infinite slabs separated by a distance of $D$. The magnitude of  $D$ must be much much larger than that of the thermal wave length of the massive bosons. Secondly, the shift of density in the excited states of the imperfect Bose gas due to the MFP, and some other characteristic properties across the phase separating point will be studied. Finally different features of the Casimir force and critical behavior will be examined with some rigorousness. We should mention here that, the reliability of the derived analytic expressions describing different properties of the imperfect Bose gas (IMBG) should be justified by comparing with other known results. To this end, we intend to examine the limiting cases as the MFP tends to zero (i.e. $a\rightarrow 0$). We note here that, in the limit $a\rightarrow 0$ all results (presented below) show complete coincidence with those of the ideal Bose gas as expected. 

The magnitude of the Casimir effect depends on the boundary conditions (bc) used. So, we intend to use the Neumann boundary condition explicitly in the theory and then to discuss the possible impact of other boundary conditions on the results. We note here that the sums for the periodic, Neumann and Dirichlet boundary conditions starts at $-\infty, 0$ and $1$, respectively. However, the latter two bcs can be expressed in terms of sums from $-\infty$ to $+\infty$. On the other hand, the operator relations for Neumann and Dirichlet bcs are $\sum_{0}^{\infty}\rightarrow \frac{1}{2}(\sum_{-\infty}^{+\infty}+1)$ and $\sum_{1}^{\infty}\rightarrow \frac{1}{2}(\sum_{-\infty}^{+\infty} -1)$, respectively. The periodic bc is idealistic in the sense that the corresponding surface energy density becomes zero, while for others it becomes nonzero. We also note that the magnitude of Casimir interaction energy is the largest for periodic bc and smallest for Dirichlet one [22].

This report is organized in the following way. Derivation of the one particle density of states in d-dimensional space, grand canonical free energy, density shift, Casimir amplitude and force are made in section 2. The critical behavior is also examined in the same section. 
Section 3 is devoted to present the results and discussions. This report is concluded in section 4.       
%\label{}
\section{Theoretical evaluation}
\subsection{One particle density of states in d-dimensional configuration space}
The repulsive pair interaction between a pair of identical massive bosons can be described within the mean field theory by $a/V$
where $a$ is a positive constant and $V$ is the volume of the Bose gas. Let us now look at a particular boson
in the $N$ boson system which is moving in the mean field due to the rest of $(N-1)$ particles. Obviously, the average potential energy experienced by the tagged boson is $\frac{a}{V}(N-1)\approx \frac{a}{V}N$. The one boson Hamiltonian
will, therefore, be
\begin{equation}
H = \frac{p^{2}}{2 m} + a n
\end{equation}
where $p$ is the momentum of the boson and $m$ is its mass; $n$ denotes the number density of the bosons ($N/V$). 

Let us assume that, the bosons are enclosed in a $d$-dimensional volume (denoted by $V^{(d)}$) $V^{(d)}=L^{d}$, $L$ being the edge of the rectangular box and
$L\rightarrow \infty$. The spacing between energy levels will therefore be very small, so the summation over
the states can be  replaced approximately by an integration. Therefore, the bulk density of states in the phase space with spatial d-dimension is
\begin{align}
\gamma(\epsilon)&=\frac{1}{(2 \pi \hbar)^{d}} \int d^{d}r \int d^{d}p \,\,\delta(\epsilon -\frac{p^{2}}{2\,m}-a\,n)\nonumber\\ &= V^{(d)}\,(\frac{m}{2 \pi\hbar})^{\frac{d}{2}}\, \frac{1}{\Gamma(\frac{d}{2})}\,(\epsilon-a\,n)^{\frac{d-2}{2}}.
\end{align}
$\epsilon$ in equation (2) is the energy eigen value of the Hamiltonian operator of equation (1).

We now assume that the Bose gas in the mean field potential (henceforth will be referred to as the imperfect Bose gas, IMBG) is placed within two infinite parallel slabs such that the $d$-dimensional volume 
$V^{(d)}=L^{d-1}\,D$. The density of states in the ($d$-1)-dimensional surface is now
\begin{align}
\gamma_{1}(\epsilon)= V^{(d-1)}\,(\frac{m}{2 \pi\hbar})^{\frac{d-1}{2}}\, \frac{1}{\Gamma(\frac{d-1}{2})}\,(\epsilon-a\,n)^{\frac{d-3}{2}}.
\end{align}
\subsection{The grand potential of the system}
As $D$ is finite, the spacing of the energy levels will be large along the $d$th direction. So, the summation over energy levels along the $d$th direction cannot be approximated by integration. Then the grand potential (with wave number $k_{d}=\frac{\pi}{D}\,l;\, l= 0,1,2,3. ....$), may be expressed applying the Neumann boundary condition as
\begin{align}
\frac{\phi_{D}(T,\mu)}{k_{B}T}=&\sum_{l=0}^{\infty}\,\int_{0}^{\infty}\,ln(1-z\,\exp(-\frac{\beta \pi^{2} \hbar^{2}}{2mD^{2}}\,l^{2})\, \exp(-\beta\,\epsilon))\,\gamma_{1}(\epsilon)\,d\epsilon\nonumber\\=& -V^{(d-1)}\,(\frac{m}{2 \pi\hbar})^{\frac{d-1}{2}}\, \frac{1}{\Gamma(\frac{d-1}{2})}\frac{1}{\beta^{\frac{d-1}{2}}}\sum_{l=0}^{\infty}\sum_{r=1}^{\infty}\frac{z^{r}}{r^{\frac{d+1}{2}}}\nonumber\\ \times &\exp(-\frac{\pi \lambda^{2}l^{2}\,r}{4 D^{2}}\, \exp(-\beta a n r)\,\Gamma(\frac{d-1}{2},-\beta a n r)
\end{align}
 where $z=\exp(\beta \mu)$,$\,\beta$ is inverse temperature times Boltzmann constant, $\Gamma(\frac{d-1}{2},-\beta a n r)$  the lower incomplete gamma function and $\lambda=h/\sqrt{2 \pi m k_{B} T}$, $h$ being the Planck's constant. Now separating the $l=0$ term and using the following identity relation
\begin{equation}
\sum_{l=1}^{\infty}e^{-\pi \alpha l^{2}}=(\frac{1}{2\sqrt{\alpha}}-\frac{1}{2})+\frac{1}{\sqrt{\alpha}}\sum_{l=1}^{\infty} e^{-\pi\,l^{2}/\alpha}
\end{equation}
one can write,
%\begin{align}
%\frac{\phi_{D}(T,\mu)}{k_{B}T}& = -V^{(d-1)} (\frac{m}{2 \pi\beta \hbar^{2}})^{\frac{d-1}{2}}\,\frac{1}{\Gamma(\frac{d-1}
%{2})}\,\,\left[\frac{1}{2}\sum_{r=1}^{\infty}
%\frac{z^{r}}{r^{\frac{d+1}{2}}} e^{-\beta a n r} \,\,\Gamma(\frac{d-1}{2},-\beta a n r)\right .\nonumber \\&+  
%\frac{D}{\lambda}\sum_{r=1}^{\infty}\frac{z^{r}}{r^{\frac{d+2}{2}}} e^{-\beta a n r}\,\, \Gamma(\frac{d-1}{2},-\beta a n r)%\nonumber\\
%& \left .+2 (\frac{D}{\lambda})\sum_{l=1}^{\infty}\sum_{r=1}^{\infty}\frac{z^{r}}{r^{\frac{d+2}{2}}}\,e^{-\beta a n r}\,\, %\Gamma(\frac{d-1}{2},\beta a n r) e^{-4 \pi (\frac{D}{\lambda})^{2}l^{2}/r}\right].
%\end{align}
\begin{align}
\frac{\phi_{D}(T,\mu)}{k_{B}T}& = -V^{(d-1)} (\frac{m}{2 \pi\beta \hbar^{2}})^{\frac{d-1}{2}}\,\,\left[\frac{1}{2}\sum_{r=1}^{\infty}\frac{z^{r}}{r^{\frac{d+1}{2}}} e^{-\beta a n r} \,\,\right .\nonumber 
\\&+  \frac{D}{\lambda}\sum_{r=1}^{\infty}\frac{z^{r}}{r^{\frac{d+2}{2}}} e^{-\beta a n r}\,\,\nonumber\\
& \left .+2 (\frac{D}{\lambda})\sum_{l=1}^{\infty}\sum_{r=1}^{\infty}\frac{z^{r}}{r^{\frac{d+2}{2}}}\,e^{-\beta a n r}\,
\,  e^{-4 \pi (\frac{D}{\lambda})^{2}l^{2}/r}\right]
\end{align}
here, we have assumed that 
\begin{equation}
\frac{\Gamma(\frac{d-1}{2},-\beta a n r)}{\Gamma(\frac{d-1}{2})}\approx 1 .
\end{equation}
Equation (7) is valid when the magnitude of lower limit of the integration in the incomplete gamma function is very small; in other words, when the strength of interaction is weak. We note that the lower limit contains a sum over $r$. So a question 
remains whether the approximation is valid for large value of $r$ or not. The answer is yes because from equation (4) we see that the lower limit 
of the incomplete gamma function, $-\beta a n r$ becomes large for large $r$ and changes the value of gamma 
function to some extent. But, at the same time the multiplying factor 
$e^{-\beta a n r}$ in  the term goes to be very small relatively. As a result, the contribution of each term to the energy at large $r$ is insignificant despite the ratio of gamma functions in Eq.(7) deviates from 1. However,
for the present study, equation (6) is the final expression for the grand potential that involves the $d$-dimensional configuration space and the mean field potential (MFP) energy.

In the absence of the  MFP energy i.e. with $a\rightarrow 0$ equation (6) stands as
\begin{align}
\frac{\phi_{D}(T,\mu)}{k_{B}T}& = -\frac{V^{(d-1)}}{\lambda^{d-1}}\,\,\left[\frac{1}{2} g_{\frac{d+1}{2}}(z) 
+  \frac{D}{\lambda} g_{\frac{d+2}{2}}(z)\right .\,\,\nonumber\\
& \left .+2 (\frac{D}{\lambda})\sum_{l=1}^{\infty}\sum_{r=1}^{\infty}\frac{z^{r}}{r^{\frac{d+2}{2}}}\,\,
\,  e^{-4 \pi (\frac{D}{\lambda})^{2}l^{2}/r}\right]
\end{align}
where $g_{\alpha}(z)$ is the Bose-Einstein function [36,37]. For $d=3$, equation (8) becomes as
\begin{align}
\frac{\phi_{D}(T,\mu)}{k_{B}T}& = -\frac{V^{(2)}}{\lambda^{2}}\,\,\left[\frac{1}{2} g_{2}(z) 
+\frac{D}{\lambda} g_{\frac{5}{2}}(z)\right .\,\,\nonumber\\
& \left .+2 (\frac{D}{\lambda})\sum_{l=1}^{\infty}\sum_{r=1}^{\infty}\frac{z^{r}}{r^{\frac{5}{2}}}\,
  e^{-4 \pi (\frac{D}{\lambda})^{2}l^{2}/r}\right .].
\end{align}
In equation (9) the bulk energy density is $-k_{B}T\,g_{\frac{5}{2}}(z)/\lambda^{3}$ and the surface energy density is $-g_{2}(z)/\lambda^{2}$ when $\mu<0$ and, these densities in the condensate ($\mu=0$) become  $-k_{B}T\,\zeta(\frac{5}{2})/\lambda^{3}$ and $-k_{B}T\zeta(2)$, respectively. The third term on the right hand side is known as the Casimir interaction term
\begin{equation}
\frac{\delta\omega_{D}}{k_{B}T} = -2 \frac{D}{\lambda^{3}}\sum_{l=1}^{\infty}\sum_{r=1}^{\infty}\,\frac{z^{r}}{r^{\frac{5}{2}}}\,e^{-4\pi (\frac{D}{\lambda})^{2}l^{2}/r}
\end{equation}
where $\delta \omega_{D}= \phi_{D}/V^{(d-1)}$. It is noticed that, equation (10) completely agrees with the properties of ideal Bose gas (IBG) [38,39]. This result, however, confirms that, the first and the second terms in the d-dimensional expression ( Equation (6)) are surface and bulk energy densities, respectively. The third term is nothing but the Casimir interaction in the d-dimensional configuration space for the IMBG.

\subsection{Critical Temperature}
The second term on the right hand side of equation (6) describes the bulk energy density
\begin{align}
\omega_{b}&=
%-\frac{k_{B} T}{\lambda^{d}}\sum_{r=1}^{\infty}\frac{e^{\beta(\mu-a n)r}}{r^{\frac{d+2}{2}}}\nonumber \\
 -\frac{k_{B}T}{\lambda^{d}}\,g_{\frac{d+2}{2}}(z^{\prime})
\end{align}
where $z^{\prime}=e^{\beta(\mu-a n)}$. From this knowledge of the bulk energy density in d-dimension one can easily evaluate the bosonic number density in the bulk,
\begin{align}
n^{(d)}&=-\frac{\partial \omega_{b}}{\partial \mu}\nonumber \\ 
%&=\frac{1}{\lambda^{d}}\,g_{\frac{d}{2}}(z^{\prime})\nonumber \\
       &= \left(\frac{\sqrt{2\pi m}}{h}\right)^{d}\,\left(k_{B}T\right)^{\frac{d}{2}}\,g_{\frac{d}{2}}(z^{\prime}). 
\end{align}
So, one can write
\begin{align}
n^{(d)}\le \left(\frac{\sqrt{2\pi m}}{h}\right)^{d}\,\left(k_{B}T\right)^{\frac{d}{2}}\,\zeta(\frac{d}{2}) 
\end{align}
where $\zeta$ denotes the Riemann zeta function. Now at the critical density one can write
\begin{equation}
n_{C}^{(d)} = \left(\frac{\sqrt{2\pi m}}{h}\right)^{d}\,\left(k_{B}T_{C}\right)^{\frac{d}{2}}\,\zeta(\frac{d}{2}) 
\end{equation}
which leads to the critical temperature in d-dimensional space
\begin{equation}
T_{C}=\left(\frac{h}{\sqrt{2\pi m}}\right)^{2}
\frac{1}{k_{B}}\left(\frac{n_{C}^{(d)}}{\zeta(\frac{d}{2})}\right)^{\frac{2}{d}}.
\end{equation}
From equation (15) one can easily show that the fraction of Bose particles in the ground state, 
\begin{equation}
\frac{n_{0}}{n}= 1-\left(\frac{T}{T_{C}}\right)^{\frac{d}{2}} 
\end{equation}
where $n$ is the sum of the number of bosons per unit volume in the ground and excited states i.e. $n=n_{0}+n^{(d)}$. 

This is of great interest to see the affect of the MFP on the shift of density of the excited bosons. In
order to examine it we consider the lowest order of MFP to have $z^{\prime}=e^{\beta(\mu-an)}\approx 
e^{\beta\mu}(1 - \beta a n)$. For the MFP to be weak $\beta a n << 1$, so it is obvious that $e^{\beta \mu}>>\beta a n\, e^{\beta \mu}$. Now expanding the Bose-Einstein function by Taylor series about $e^{\beta\mu}$ we have
the shift of density of the excited bosons
\begin{equation}
\frac{\Delta n^{(d)}}{n}=-(\frac{\sqrt{2 \pi m}}{h})^{d} \,a\, (k_{B}T)^{\frac{d}{2}-1} \,g_{\frac{d}{2}-1}(e^{\beta \mu})
\end{equation}
It is seen from equation (17) that the shift of the density vanishes as $a\rightarrow 0$ or $T\rightarrow 0$, which justifies, at least qualitatively, that our derivation is correct. 
\subsection{Casimir amplitude}
For $D/\lambda>>1$ the summation over $r$ in equation (6) may be approximated by an integration. This approximation leads to 
\begin{equation}
\frac{\delta \omega_{D}}{k_{B}T}=\frac{\Delta (d,\sigma)}{D^{d-1}}
\end{equation}
where the Casimir amplitude in units of $k_{B}T$ is
\begin{equation}
\Delta(d,\sigma)= -\frac{1}{2^{d-2} \pi^{\frac{d}{2}}}\,\sum_{l=1}^{\infty} (\frac{\sigma}{l})^{\frac{d}{2}}\,K_{\frac{d}{2}}(2\,\sigma\,l)
\end{equation}
with
\begin{equation}
\sigma = 2 \sqrt{\pi}\,(\frac{D}{\lambda})(\beta a n-\beta \mu)^{\frac{1}{2}},\nonumber 
\end{equation}
where $K_{\alpha}$ denotes the modified Bessel function. $\Delta(T,\sigma)$ in equation (18) gives a non-zero magnitude
at the critical point and depends only on the boundary condition for a specific dimension. Away from the critical regime it decays exponentially 
(exponential form is implicit in $K_{\frac{d}{2}}(\alpha)$) and thus becomes zero for large $D$. So, $\Delta(T,\sigma)$ can be referred to as the Casimir amplitude in a way analogous to the ideal Bose gas[38].
As the IMBG approaches to the point of phase transition i.e. $\mu\rightarrow a n_{c}^{(d)}$, $\sigma$ becomes very small. In this situation equation (19) reduces to 
\begin{equation}
\Delta(d)= -\frac{1}{2^{d-2} \pi^{\frac{d}{2}}}\,\Gamma(\frac{d}{2})\,\zeta(d) .
\end{equation}
Here, we have used the relation $K_{\alpha}(x)\approx \frac{\Gamma{(\alpha)}}{2}\,(\frac{2}{x})^{\alpha}$ as $x\rightarrow 0$. For $d=3$, the Casimir amplitude stands as $\Delta= -\zeta(3)/(8 \pi)$ which coincides exactly with the result of the ideal Bose gas [38].

\subsection{Casimir force}
The Casimir force corresponding to equation (6) is
\begin{equation}
F_{C} = -\frac{\partial}{\partial D}(\delta \omega_{D})= -\frac{k_{B} T}{2^{\frac{3d-4}{2}}\,\pi^{\frac{d}{2}}}\,\sum_{l=1}^{\infty}\frac{(2 \sigma l)^{\frac{d+2}{2}}\,K_{\frac{d+2}{2}}(2 \sigma l) - (2\sigma l)^{\frac{d}{2}}\,K_{\frac{d}{2}}(2 \sigma l)}{l^{d} D^{d}}
\end{equation}
Now for the IBG, $a=0$ and then consideration of $\mu\rightarrow 0$ in equation (21) gives
\begin{equation}
F_{C} = -\frac{k_{B} T}{2^{\frac{3d-4}{2}}\,\pi^{\frac{d}{2}}}\,\frac{[\Gamma(\frac{d+2}{2}) 2^{\frac{d}{2}} - \Gamma(\frac{d}{2})\,2^{\frac{d}{2}-1}]}{D^{d}}\,\zeta(d).
\end{equation}
If we now examine this expression for $d=3$, we have
\begin{equation}
F_{C} = -\frac{k_{B} T}{4\pi}\,\frac{\zeta(3)}{D^{3}}.
\end{equation} 

In order to reexamine the result found in equation (23) and, also to see explicitly the behavior of exponential decay we first integrate the third term on the right side of equation (6) over $r$ in the three dimensional space. Secondly, we differentiate the resulted expression with respect to $D$. We finally arrive at the following expression for the Casimir force 
\begin{equation}
F_{C} = -\frac{k_{B} T}{8 \pi}\,\left[\sum_{l=1}^{\infty}(\frac{2}{D^{3}}+\frac{2 l}{D^{2} \eta})\,\frac{1}{l^{3}}(1+2 l \frac{D}{\eta})
-\frac{2}{D^{2} l^{2} \eta}\right]\,e^{-2 l\frac{D}{\eta}}.
\end{equation}
where the characteristic decay length 
\begin{equation}
\eta = \frac{0.5\,\lambda}{[\beta \pi (an-\mu)]^{\frac{1}{2}}}.
\end{equation}
In the condensate ($T\leq T_{C}$) $\mu=\mu_{c}=an_{c}$. In this case $\eta\rightarrow\infty$ and equation (24) reduces to the form of equation (23),
\begin{equation}
F_{C} = -\frac{k_{B} T}{4\pi}\,\frac{\zeta(3)}{D^{3}}.
\end{equation} 
Again for $T> T_{C}$, $\mu \neq \mu_{C}$, as a result $\eta$ is finite. Then, due to the fact that $D/\lambda>>1$,
 $\frac{D}{\eta} >>1$, so $F_{C}\sim e^{-2 D/\eta}$. The Casimir force, therefore, shows a power 
law behavior in the condensate, and decays exponentially in the non-condensate (i.e.for $T>T_{C}$) 
with a characteristic length $\eta$. It is also noticed from equation (21) that for $T<T_{C}$ the Casimir
 force decreases with decreasing value of T and vanishes at $T=0$ K.

\subsection{Critical exponents near the point of Bose condensation}
In order to examine the values of the critical exponent let us start with the characteristic decay length for a IBG for which $a=0$. In this case equation (25) yields
\begin{equation}
\eta \sim (-\mu)^{-\frac{1}{2}}. 
\end{equation}
The critical exponent is therefore $1/2$. This is the same value as shown by others [33,39,40]. The critical exponent in the case of d-dimensional space may be obtained in the following way.
When $\alpha \rightarrow 0$, one can easily show for $2<d<4$ that [1]
\begin{equation}
g_{\frac{d}{2}}(e^{\alpha})=\zeta(\frac{d}{2})+\Gamma(1-\frac{d}{2})\,\alpha^{\frac{d-2}{2}}
\end{equation}
where $\alpha=\beta(\mu-an)$. A straight forward manipulation leads to (with $\lambda^{d}\mu_{c}= a\zeta(d)$ and $\lambda^{d}\mu= a\,g_{\frac{d}{2}}$)
\begin{equation}
\beta(\mu-an)=\left[\lambda^{d}(\mu-\mu_{c})\,\frac{1}{a\Gamma(1-\frac{d}{2})}\right]^{\frac{2}{d-2}}.
\end{equation}  
Now substituting equation (29) into (25) we have the characteristic decay length in the $d$-dimensional configuration space
\begin{equation}
\eta = \frac{0.5\,\lambda}{[\frac{a\Gamma(1-\frac{d}{2})}{\lambda^{d}}(\mu-\mu_{c})]^{\frac{1}{d-2}}}
\sim (\mu-\mu_{c})^{-\frac{1}{d-2}}
\end{equation} 
So, for the IMBG in d-dimensional space the critical exponent is 1/(d-2). For d=3 it is just 1 unlike 
ideal Bose gas for which it is 1/2. The calculated magnitude of the exponent agrees well with values evaluated by using a different theoretical method [26].
\section{Results and Discussions}
Properties of massive boson gas in the  presence  of the mean field potential, and confined between two parallel slabs are presented in this section. This study  is performed by using the density of states  approach with spatial dimension $d$. Most importantly, this approach allows us to study properties of the two phase system, such as the condensed and normal phase across the Bose condensation point and, also to examine the critical exponents close to the point of phase transition.

Equation (15) shows the critical temperature, $T_{C}$, for the Bose condensation of the IMBG. Here  $T_{C}$ is proportional to a factor $[n_{c}^{(d)}/\zeta(\frac{d}{2})]^{\frac{2}{d}}$ which is valid for $d>2$ because at $d=2$ the Zeta function becomes infinite and consequently $T_{C}= 0$. On the other hand, $n_{C}^{(d)}$ diverges for $d=2$. Equation (15), however, demonstrates that the critical temperature decreases with increasing $d$, but for $d\rightarrow \infty$ the critical temperature becomes a constant, $T_{C}=(h/\sqrt{2 \pi m})^{2}/k_{B}$.

It is noticed from equation (16) that the fraction of the boson number in the condensate varies with the dimensionality in such a way that $n_{0}/n$ increases non-linearly with increasing power of $d$ and,  
 as $d\rightarrow \infty$ all bosons drop to the ground state for $0<T<T_{C}$. Equation (16) also shows that at $T=T_{C}$ no particles are there in the ground state and, at $T=0$ all bosons fall to it. But in the temperature range  $0<T<T_{C}$ and for finite $d$ some bosons remain in the excited state.  Equation (17)
shows the shift of density in the lowest order due to the presence of MFP. The sign of the shift is negative, that means the density of the excited bosons reduces due to the repulsive mean field potential.  It is seen from equation (17) that the shift of the density vanishes as $a\rightarrow 0$, that means if there is no field there is no density shift. It is also noticed that as $T\rightarrow 0$ shift of density becomes zero. The reason of it is the following. At $T=0$ K all bosons in the condensed state fall to the ground state as we have described before. So no particle remains in the excited state to cause the shift to happen.  

Figure 1 shows the Casimir amplitude (in unit of $k_{B}T$) for $\mu=\mu_{c}$ (Eqn. (20)) and for $ \mu \neq \mu_{c}$ ( Eqn. (19)). The former denotes the condensed state and the latter the normal state. But both properties approach to zero with increasing dimension $d$. Figure illustrates that, in the condensed state (i.e. $\mu=\mu_{c}$) the values of $\Delta(d,\sigma)$ is negative and large at lower dimension and increases roughly exponentially for the large value of $d$. In the case of non-condensed state the values of the Casimir amplitude are positive and large at lower dimension, becomes negative at around $d=3$ and, finally approach to zero with further increment of $d$. In the condensed state($\mu=\mu_{c}$ ) the Casimir amplitude is $\Delta(3)= -\zeta(3)/(8\,\pi)$, $\zeta(3)$  being the Riemann zeta function. This magnitude is just equal to that of the ideal Bose gas [38]. We note here that, besides dimensionality, the magnitude of the Casimir amplitude depends on the boundary condition employed. For example, $\Delta(3)$ in the case of periodic boundary condition is just eight times larger than that of the Neumann or Dirichlet boundary condition [33]. However, equation (20) clearly states that $\Delta(3)$ in condensate is not directly affected by the MFP. The reason is that, in the condensate $\mu_{c} = a n_{c}$ which makes $z^{\prime}=1$. Consequently, the Casimir amplitude becomes independent of MFP.
 \begin{figure}
 \includegraphics[width=5cm,height=6cm]{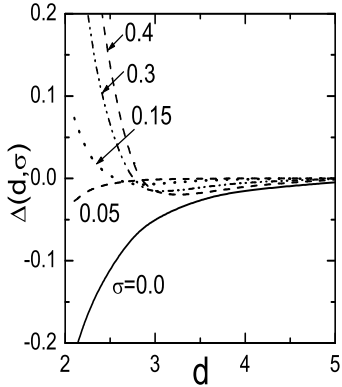}
 \caption{d-dependence of the Casimir amplitude, $\Delta(d,\sigma)$; $\sigma$ is defined in the text.}
 \end{figure}

Equation (21) expresses the Casimir force in the case of a IMBG confined between two infinite slabs placed in a $d$-dimensional configurational space. In the limit $a\rightarrow 0$ (i.e. for an ideal Bose gas) this equation reduces to the form of
equation (22) which finally takes up the shape of equation (23) for $d=3$. This completely agrees with the previously established expression for the IBG [14]. This, in turn, justifies that equation (21) correctly 
represents the Casimir force between two slabs in $d$-dimensional configuration space. The modified Bessel function $K_{\frac{d}{2}}$ in
equation (21) decays exponentially, consequently the Casimir force decays in the same manner. This decay property can be demonstrated more spectacularly by evaluating the Casimir force in a spacial case for $d=3$ (see equation (24)). Here, equation (24) illustrates that the decay is happening with a characteristic decay length $\eta$. We note that the magnitude of $\eta$ for a IMBG depends on the boundary condition used, for example,  $\eta_{N}=\eta_{D}=\eta_{P}/2$ where subscript N, D and P denote the Neumann, Dirichlet and periodic boundary conditions, respectively [40]. We also note that the decay length is related to the bulk correlation length as $\xi\rightarrow \sqrt{\pi}\eta$ in the case of IMBG and, as $\xi\rightarrow \eta_{P}$ in the case of non-condensed phase.
This equation also shows that the exponential decay of the Casimir force is also related to the separation distance, $D$,  of the two slabs and goes to zero when $D$ is large. It is worth mentioning, here, that as $\mu \rightarrow \mu_{c}$ equation (24) yields the same expression for $F_{C}$ as equation (23). This result also agrees with that of IBG [14]. 
%%Figure 2 illustrates the logarithmic value of the Casimir force  -$log|F_{c}|$ as a function of dimension $d$ for an 
%arbitrary value of temperature and $D$. -$log|F_{c}|$ decreases linearly with increasing $d$ in both condensed and non-
%condensed phases near the condensation point.  
%\begin{figure}
% \includegraphics{Casimir_Force.EPS}
% \caption{d-dependence of the Casimir force. Force is calculated with $T=0.1\,$K, $D=10^{-6}$ m.}
% \end{figure}

Finally, we turn to the results of critical exponents evaluated near the Bose condensation point. Equation (27) shows that for an ideal Bose gas the critical exponent is $1/2$. On the other hand, equation (30) shows that, in the case of IMBG the exponent is $1/(d-2)$ in the $d$-dimensional configuration space which reduces
to 1 for $d=3$. Most importantly, the value of the critical exponent varies as an inverse function of dimensionality $d$. This result agrees with the findings of others [26,33] although they used a different theoretical approach. 
\section{Conclusion}
The Casimir interaction energy for an IMBG confined between two parallel infinite slabs separated by a distance $D$ in a particular direction in a $d$-dimensional space is evaluated, in this report, by using the one particle density of states method. Here the Neumann boundary condition is applied, unlike other authors, and the implication of other boundary conditions are discussed where appropriate. This study nicely describes the amount of density shifts in the lowest order due to the presence of MFP and, to the best of our knowledge this is done for the first time. This shift disappears while the potential energy goes to zero that means no shift happens in the absensce of a potential. The Casimir force obtained by taking derivative of the surface interaction term is found to  vary as an inverse power law of separation distance $D$ in the condensate and, to decay exponentially in the normal state close to the condensation point. The exponent in the power law is solely determined by the dimensionality $d$ ($\sim D^{-d}$). The Casimir force decreases with decreasing value of temperature when $T<T_{C}$ and vanishes at $T=0$. On the other hand, in the high temperature limit i.e. for $T>>T_{C}$ the exponential factor of the Casimir force approaches to zero. Physically this means that, for $T>>T_{C}$ the  Bose-Einstein statistics approaches to the Maxwell- Boltzmann one, consequently the Casimir interaction term arising due to quantum effect does not appear in the grand potential and thus the Casimir force ceases to exist. The critical exponent in the case of the IMBG is $1/(d-2)$ which reduces to $1$ in the 3-dimensional configuration space whereas the value of the exponent in the case of IBG is just $1/2$. Regarding the critical exponent, equation (30) is valid for any value of $d$ except for those of even integers. We finally conclude with
the remark that, in case of IMBG, the one particle density of states approach is capable to predict the same results as those predicted by the rigorous N-body partition function method [33,40] in a rather simpler way. Moreover, the present approach correctly describes the amount of density shift which yet to be seen from any rigorous approach. 
\section{Acknowledgement}
I am thankful to the EMMA-2 program for financial support. Thanks to Professor Marek Napiorkowski for his critical comments in a discussion with him. Thanks also go to Mir Mehedi Faruk for reading the manuscript and for showing some typographic mistakes therein.\\\\    
%\bibliography{your-bib-file}
%\begin{thebibliography}{99}
References\\\\
$[1]$ R. M. Ziff, G. E. Uhlenbeck and M. Kac, Phys. Reports 32, 169 (1977).\\
$[2]$ R. Beckmann, F. Karsch and D. E. Miller, Phys. Rev. Lett. 43, 1277(1979).\\
$[3]$ S. Fujita, T Kimura and Y. Zheng, Foundation of Phys. 21,1117(1991).\\
$[4]$ K Huang, Statistical Mechanics, John Wiley and Sons, Inc. New York 1963.\\
$[5]$ L. J. Landau and I. F. Wilde, Commun. Math. Phys. 70, 43 (1979).\\
$[6]$ M. H. Anderson, J. R. Ensher, M. R. Matthews, C. E.Wieman and E. A.Cornell, Science 269, 198(1995).\\
$[7]$ C. C. Bradley, C. A. Sackett, J. J. Tollett and R. G. Hulet, Phys. Rev. Lett. 75, 1687 (1995).\\
$[8]$ K. B. Davis, M.-O. Mewes, M. R. Andrews, N. J. van Druten, D. S. Durfee, D. M. Kurn and W. Ketterle, 
           Phys. Rev.   Lett. 75, 3969 (1995).\\
$[9]$ L. Slasnich, J. Math. Phys. 41,8016(2000).\\
$[10]$ S. Biswas, Eur. Phys. J. D 42,109(2007).\\
$[11]$ T. T. Chou, C. N. Yang and L. H. Yu, Phys. Rev. A53, 4257(1996).\\
$[12]$ E. B. Davies, Commun. Math. Phys. 28, 69(1972).\\
$[13]$ D. W. Robinson, Commun. Math. Phys. 50,53 (1976).\\
$[14]$ T. Lin, G. Su, Q. A. Wang and J. Chen, Euro. Phys. Lett. 98, 40010 (2012).\\
$[15]$ P. A. Martin and J. Piasecki, Phys. Rev. E 68, 016113 (2003).\\
$[16]$ J. R. Ensher, D. S. Jin, M. R. Matthews, C. E. Wieman and E. A. Cornell, Phys. Rev. Lett. 77, 4984 (1996).\\
$[17]$ V. I. Yukalov and E. P. Yukalova, Laser Phys. Lett. 2, 506 (2005).\\
$[18]$ V. I. Yukalov and H. Kleinert, Phys. Rev. A 73, 063612 (2006).\\
$[19]$ M. Wilkens, F. Illuminati and M. Kr$\ddot{a}$mer, J. Phys. B: Mol. Opt. Phys. 33, L779 (2000).\\
$[20]$ Gordon Baym, Jean-Paul Blaizot, Markus Holzmann, Franck Lalo$\ddot{e}$, and D. Vautherin, Phys. Rev. 
      Lett. 83, 1703  (1999).\\
$[21]$ A. Griffin, Phys. Rev. B 53, 9341 (1996).\\
$[22]$ Ariel Edery, J. Phys. A: Math. Gen. 39, 685 (2006).\\
$[23]$ J. Schiefele and C. Henkel, J. Phys. A: Math. Theor. 42, 045401 (2009).\\
$[24]$ Ariel Edery, J. Stat. Mech. 2006, P06007 (2006).\\
$[25]$ M. Li, L. Chen, and C. Chen, Phys. Rev. A 59, 3109(1999).\\
$[26]$ M. Napiorkowski, P. Jakubczyk and K. Nowak, J. Stat. Mech. 2013, P06015 (2013).\\
$[27]$ P. Jakubczyk and N. Napiorkowski, J. Stat. Mech. 2013, P10019 (2013).\\
$[28]$ A. Jellal and M. Daoud, Mod. Phys. Lett. B 17, 1321 (2003).\\
$[29]$ Xiao-Lu Yu, Ran Qi, Z. B. Li and W. M. Liu, Euro. Phys. Lett. 85, 10005 (2009).\\
$[30]$ P. Jakubczyk and M. Napiorkowski, Phys. Rev. B 87, 165439(2013).\\ 
$[31]$ A. B. Acharyya and M. Acharyya, Acta Phys. Polonica B 43, 1805 (2012).\\
$[32]$ A. Gambassi, J. Phys.: Conf. series 161, 012037 (2009).\\
$[33]$ M. Napiorkowski and J. Piasecki, Phys. Rev. E 84, 061105 (2011).\\
$[34]$ M. van der Berg, J. T. Lewis and P. de Smedt, J. Stat. Phys. 37, 697(1984).\\
$[35]$ J. V. Pule and V. A. Zagrebnov, J. Phys. A: Math. Gen 37, 8929 (2004).\\
$[36]$ M. Kardar, Statistical Physics of Particles, Cambridge University Press, Cambridge, 2007.\\
$[37]$ R. K. Pathria and P. D. Beale, Statistical Mechanics, 3rd ed. Academic Press, Oxford, UK (2007).\\
$[38]$ P. A. Martin and V. A. Zagrebnov, Europhys. Lett. 73, 15 (2006).\\
$[39]$ A. Gambassi and S. Dietrich, Europhys. Lett. 74, 754 (2006).\\
$[40]$ M. Napiorkowski and J. Piasecki, J. Stat. Phys. 147, 1145 (2012).
%\end{thebibliography}
\end{document}